\documentclass[12pt]{iopart}
\usepackage{bm}
\usepackage{url}
\usepackage{graphicx}
\usepackage{xcolor}
\usepackage{hyperref}
\usepackage{hhline}

\expandafter\let\csname equation*\endcsname\relax
\expandafter\let\csname endequation*\endcsname\relax
\usepackage{mathtools}
\usepackage{empheq}

\usepackage{subcaption}

\DeclareFontFamily{OT1}{pzc}{}
\DeclareFontShape{OT1}{pzc}{m}{it}{<-> s * [1.10] pzcmi7t}{}
\DeclareMathAlphabet{\mathpzc}{OT1}{pzc}{m}{it}

\begin{document}
\title[Optimal collision avoidance in swarms of active Brownian particles]{Optimal collision avoidance in swarms of active Brownian particles}

\author{Francesco Borra} \address{Dipartimento di Fisica, Universit\`a
  ``Sapienza'' Piazzale A. Moro 5, I-00185 Rome, Italy}
\ead{frr.borra@gmail.com}

\author{Massimo Cencini} \address{Istituto dei Sistemi Complessi, CNR, via dei Taurini 19, 00185 Rome, Italy and INFN ``Tor  Vergata''}
\ead{massimo.cencini@cnr.it}

\author{Antonio Celani} \address{Quantitative Life Sciences, The Abdus
  Salam International Centre for Theoretical Physics–ICTP, Trieste
  34151, Italy} \ead{celani@ictp.it}

\begin{abstract}

  The effectiveness of collective navigation of biological or artificial agents requires to accommodate for contrasting requirements, such as staying in a group while avoiding close encounters and at the same time limiting the energy expenditure for manoeuvring.  Here, we address this problem by considering a system of active Brownian particles in a finite two-dimensional domain and ask what is the control that realizes the optimal tradeoff between collision avoidance and control expenditure.  We couch this problem in the language of optimal stochastic control theory and by means of a mean-field game approach we derive an analytic mean-field solution, characterized by a second-order phase transition in the alignment order parameter. We find that a mean-field version of a classical model for collective motion based on alignment interactions (Vicsek model) performs remarkably close to the optimal control. Our results substantiate the view that observed group behaviors may be explained as the result of optimizing multiple objectives and offer a theoretical ground for biomimetic algorithms used for artificial agents.

\end{abstract}

\maketitle

\section{Introduction}
Awe-inspiring examples of organized collective motions abound in a
number of biological problems
\cite{okubo1986dynamical,vicsek2012collective} from simple
microorganisms such as bacteria \cite{wolgemuth2008collective}, to
insects and higher animals which display deliberate social behaviors
such as insect swarming \cite{sullivan1981insect,cavagna2017dynamic}
bird flocking \cite{cavagna2018physics,attanasi2014information}, and
fish schooling or shoaling
\cite{pitcher1983heuristic,pavlov2000patterns}.  Most impressively,
large and dense groups of animals can organize themselves in complex
coordinated motions avoiding collisions while flying or swimming at
close distance.  Several models have been proposed to model the origin
of such phenomena in terms of simple behavioral rules.  A first
intuition of the basic ingredients came from computer graphics
\cite{reynolds1987flocks} and entered the domain of statistical
physics with the Vicsek model \cite{vicsek1995novel}: the core idea is
that each animal in the group needs to align its heading direction
with the mean direction of its neighbors. If such local alignment
interactions are strong enough with respect to the unavoidable noise
(in our case, on the heading directions) a transition from a
disordered phase to a collective-order phase, characterized by global
orientational order (alignment) of the heading directions, may occur.
This idea was also applied, for instance, to the control of groups of
artificial agents such as robots, where collision avoidance is crucial
\cite{turgut2008self,viragh2014flocking}.  Overall, these approaches
generally build agent-based models aimed at generating certain
collective behaviors starting from intuition, observations or by
reverse-engineering natural phenomena via data analysis; this often
leads to biomimetic algorithms for robotics.

In this paper, we try to approach the problem of collision avoidance from a
different perspective. We do not aim at modeling the interactions that underlie
a certain collective behavior, but instead we consider a
simple model of swarming agents and explicitly set the goal of avoiding
collisions in the form a cost function and ask the following
questions: What is the optimal choice of control which minimizes the
collective cost? How does the optimal control compare with known
agent-based models which lead to collision avoidance?

The natural setting to answer the above questions is that of optimal
control theory~\cite{todorov2009efficient} and mean-field game
formalism~\cite{lasry2007mean,ullmo2019quadratic}. Similar approaches,
indeed, have already been shown to yield promising results. For
instance, for collective search problems  the optimal control
reduces in some limit to a well known model of
chemotaxis~\cite{pezzotta2018chemotaxis}, for the problem swarming
agents in one-dimensional disordered
environments~\cite{hongler2020mean}, and also for flocking problems
with the aid of reinforcement learning
techniques~\cite{durve2020learning}.  

In what follows, we start by introducing the setting of the problem in
Sec.~\ref{sec:CMAOCP} where also the optimal control formalism is
presented. We model agents as active Brownian particles
\cite{romanczuk2012active,bechinger2016active} which move in
two-dimensions and whose heading direction is subject to rotational
noise. They try to avoid collisions by exerting some control on
their heading direction, in the form of a torque.  Collisions lead to
a cost, but also the control itself is not free of charge, and for
that we assume a quadratic dependence in the angular velocity. The
cost for control can either be understood in terms of power
dissipation, physical limitations of an animal/robot, or in terms of
the cognitive cost of deviating from free spontaneous behavior
\cite{todorov2009efficient}. Therefore, an agent is interested in
applying a non-trivial control to its motion only to the extent to
which the gain outweighs the cost: the optimal strategy emerges from
this tradeoff. This cost minimization problem can be exactly mapped
into a quantum many-body problem which is unfortunately hard to solve
in general. For this reason, we introduce a mean field approximation
(Sec.~\ref{sec:MFA}) that reduces the many-body problem to a quantum
pendulum, which is exactly solvable. Under this approximations, agents
are assumed to be homogeneously distributed. While this is unrealistic
under many respects, it can suitably describe the optimal behavior over an approximately
uniform region in the bulk of a swarm.  We show that all relevant
parameters combine into a fundamental tradeoff parameter $h$, which
effectively accounts for the balance between the collision and control
costs.  Upon increasing such tradeoff parameter, the system displays a
second order phase transition in terms of the polar order parameter (a
measure of the mean-field alignment) at a certain critical value
$h_c$. Then, we study some relevant observables, such as the cost, the
polar order and the susceptivity -- which is known to be important in
collective motions \cite{mora2011biological} -- both near the critical
point (Sec.~\ref{sec:CB}) and in the strong coupling regime (large
$h$, corresponding to collisions costs dominating, see
Sec.~\ref{sec:SC}).

Remarkably, in both limits, the optimal control is well approximated
by a sinusoidal function of the difference between the individual
heading direction (angle) and the mean one
(Sec.~\ref{sec:motsin}). Interestingly, the sinusoidal control is a
distinctive trait of the well-known Vicsek-like models
\cite{vicsek2012collective,farrell2012pattern,chepizhko2021revisiting}
and its mean-field versions
\cite{peruani2008mean,chepizhko2009kinetic,chepizhko2010relation}. This
observation motivates a \textit{vis \`a vis} comparison between the
optimal and sinusoidal control. For a sound comparison, we first find
the sinusoidal control which minimizes the total cost
(Sec.~\ref{sec:BSM}). Remarkably, such best sinusoidal model is
controlled by the same tradeoff parameter, and the polar order turns
out to display a second order transition at the same critical point as
for the optimal model; with analytical tools, we explore this regime
along with the strong coupling one. Finally, we proceed with a
systematic comparison (Sec.~\ref{sec:CBOSABSM}) in the whole range of
the tradeoff parameter, showing how the optimal solution, while close
to its sinusoidal approximation, can better manage the collisions, and
that the sinusoidal model becomes the exact optimum in the strong
coupling (large $h$) limit. The last section is dedicated to
discussion and outlook (Sec.~\ref{sec:end}).  Some derivations and
technical material are moved to the Appendices.

\section{Optimal solution of the collision problem}

\subsection{Collision minimization as an optimal control problem}~\label{sec:CMAOCP}
\begin{figure*}[b!]
\centering
\includegraphics[width=0.6\linewidth]{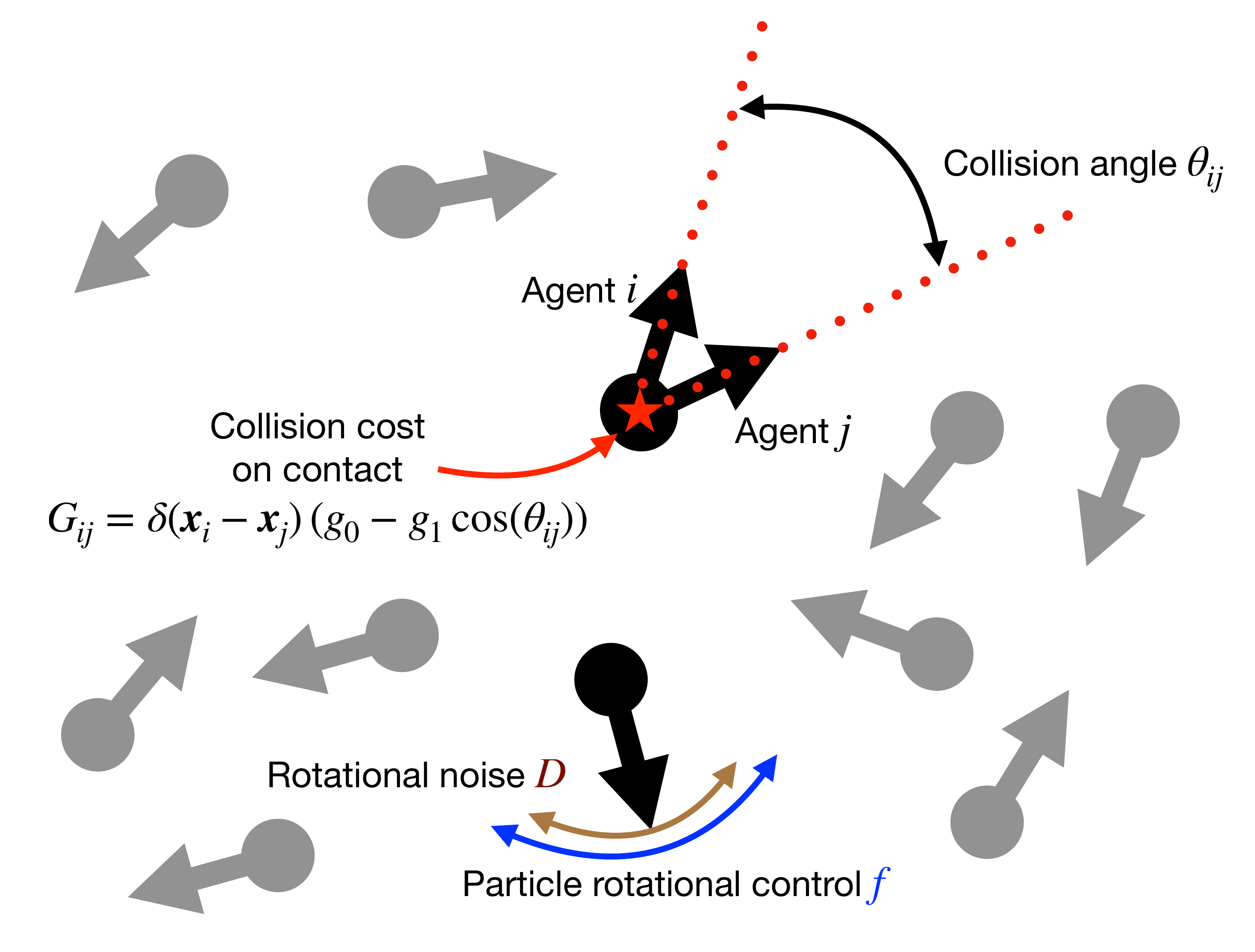}
 \caption{\label{fig:1} Sketch of the swarming active Brownian
   particles. The black particle on the bottom illustrates the angular
   dynamics influenced by rotational noise (brown arrows) and the
   control (blue arrows). The couple of black particles on the top,
   denoted $i$ and $j$, illustrates the cost paid for each contact,
   depending on the collision angle. The agents can avoid the collisions by controlling their angular
   velocity, but pay a cost for it.  }
\end{figure*}

We consider a group of $N$ agents in two dimensions whose goal is to
swarm together while avoiding collisions with each other (see
Fig.~\ref{fig:1}).  We model the agents as active Brownian
particles~\cite{romanczuk2012active,bechinger2016active}: each agent
$i$ is a self-propelled particle moving with a constant speed $u_0$ in
a direction identified by an angle $\theta_i$ (or, equivalently, by
the associated unitary vector
$\bm{n}(\theta_i)=(\cos\theta_i,\sin\theta_i)$) which randomly changes
due to rotational noise with diffusivity $D$. Each agent, to avoid
collisions, can exert some control, $f_i$, on its angular velocity and
possibly contrast the rotational noise.  The controls $f_i$ are, in
the most general case, functions of all positions, $\bm x_j$, and
heading directions, $\theta_j$, of all agents ($j=1,\ldots, N$). The
dynamics of agent $i$ thus reads
\begin{equation}\label{dyn}
\begin{dcases}
d\bm{x}_i=u_0\,\bm{n}(\theta_i) dt\\
d\theta_i=f_i(\bm{x}_i,\theta_i;\{\bm{x}_j,\theta_j\}_{j\neq i})\,dt+\sqrt{2\,D}\,d\xi_i\,,
\end{dcases}
\end{equation}
where the noise term in the angular dynamics is a zero mean, $\langle
d\xi_i(t) \rangle=0$, Gaussian process with correlation $\langle
d\xi_i(t)\;d\xi_j(t') \rangle=\delta_{ij}\,\delta(t-t')\;dt$. We 
assume periodic boundary conditions, since we are only interested in
the bulk interactions within the swarm. This choice will not be
relevant for the rest of the paper.

When particle pairs, say $i$ and $j$, collide they pay a cost
$G_{ij}=\delta(\bm{x}_i-\bm{x}_j)\,\mathcal{G}(\theta_{ij})$ with
$\mathcal{G}(\theta_{ij})$ representing the functional dependence of
cost on the collision angle, $\theta_{ij}=(\theta_i-\theta_j)$. By
expanding $\mathcal{G}(\theta)=g_0+g_1\cos\theta+g_2
\cos(2\theta)+...$ into (even) harmonics and truncating after the
second term, we obtain
$G_{ij}=\delta(\bm{x}_i-\bm{x}_j)\,(g_0-g_1\,\bm{n}(\theta_i)\cdot\bm{n}(\theta_j))=
\delta(\bm{x}_i-\bm{x}_j)\,(g_0-g_1\cos\theta_{ij})$. As we will see,
the cost per contact $g_0>0$ will be somehow unimportant but its
relationship with angular cost $g_1>0$ allows for different model
interpretations. For instance, if $g_0=0$ we have a pure alignment
problem, while for $g_1=g_0$ we have a pure collision-based model, as
in the latter case the collision cost is proportional to the relative
velocity which is exactly zero when velocities are aligned.

Agent $i$ can partially control its heading direction by
imparting an angular velocity $f_i$ but it pays a cost
$\alpha\,f_i^2/2$. The quadratic choice for the cost of control,
besides being quite natural when interpreted in terms of power
dissipation, has an information-theoretical foundation as  the cost (measured in terms of the Kullback-Leibler
divergence) of deviating from a random control strategy
\cite{todorov2009efficient}. The total cost per unit time - the sum of individual costs - reads
\begin{equation}\label{cost0}
C(\bm{x}_1,\theta_1;...;\bm{x}_N,\theta_N)= \frac{\alpha}{2}\sum_i f_i^2+\frac{1}{2}\sum_{i\neq j}\delta(\bm{x}_i-\bm{x}_j)\,(g_0-g_1\,\bm{n}(\theta_i)\cdot\bm{n}(\theta_j))\,.
\end{equation}
Notice that the parameter $\alpha>0$ can be reabsorbed in the definition of $g_0$ and $g_1$, since we are only interested in the optimal strategy, while it would have played a role in risk-sensitive scenarios~\cite{pezzotta2018chemotaxis,howard1972risk,dvijotham2011unified}.

The agents collective goal is to choose the controls that minimize
the average total cost $\bar C= \int \prod_{k=1}^N
d\bm{x}_k\,d\theta_k\;C(\bm{x}_1,\theta_1;...;\bm{x}_N,\theta_N)\;P(\bm{x}_1,\theta_1;...;\bm{x}_N,\theta_N)$, with
$P$ being the stationary joint probability density of particles
positions and angles. The non-trivial point is that $P$ itself depends
on the controls $\{f_i\}_{i=1}^{N}$ and should be determined as part
of the solution. In particular, the joint probability $P$, besides the normalization constraint  $\int \prod_{k=1}^N
d\bm{x}_k\,d\theta_k\;P(\bm{x}_1,\theta_1;...;\bm{x}_N,\theta_N) =1$,
must be the stationary solution of the
Fokker-Planck equation associated to Eq.~\eqref{dyn}, which reads
\begin{equation}\label{FP0}
\sum_{i=1}^N\left[ -u_0\,\partial_{\bm{x}_i}\;\bm{n}(\theta_i)-\sum_{i}\partial_{\theta_i}\;f_i+D\;\sum_i\partial^2_{\theta_i}\right]P=\sum_{i=1}^N\mathcal{L}_i \,P=\mathcal{L}_{(N)}P=0\,,
\end{equation}
where $\mathcal{L}_i$ is the single-agent linear Fokker-Planck operator and $\mathcal{L}_{(N)}=\sum_i\mathcal{L}_{i}$ the $N$-bodies one. By minimizing the total cost, we are looking for a cooperative solution to the problem. The solution of the constrained minimization can obtained by a generalized Lagrange-multipliers technique, or namely, by finding the stationary points of the auxiliary functional\footnote{The minus signs in Eq.~\eqref{functional} are chosen for the convenience of notation.} (Pontryagin principle~\cite{pontryagin2018mathematical})
\begin{equation}\label{functional}
  \mathcal{H}=\lambda+\int  \prod_{i=1}^N d\bm{x}_i\,d\theta_i\;\, [C-\lambda-\Phi\,\mathcal{L}_N]\,P\,.
\end{equation}
The normalization and dynamical constraints are obtained by imposing
stationarity w.r.t. (with respect to) the multipliers $\lambda$ and
$\Phi(\bm{x}_1,\theta_1;...;\bm{x}_N,\theta_N)$,
respectively.\footnote{Notice that $\Phi$  is a function because
  $\mathcal{L}_N P=0$ must be imposed for all angles and positions.} 
  The non-trivial
results come from the request of stationarity w.r.t. $P$ and $f_i$,
which yields
\begin{empheq}[left=\empheqlbrace]{align} 
  &\frac{\delta \mathcal{H}}{\delta f_i}=0\implies
  f_i=\partial_{\theta_i}\Phi \label{contrgeneral}\\ &\frac{\delta
    \mathcal{H}}{\delta P}=0 \implies
  C-\lambda-u_0\,\sum_{i}\bm{n}(\theta_i)\cdot\partial_{\bm{x}_i}\;\Phi-\sum_{i}f_i\,\partial_{\theta_i}\;\Phi-D\;\sum_i\partial^2_{\theta_i}\Phi=0 \,. \label{HJB}
    \end{empheq}
Equation~\eqref{HJB} is the Hamilton-Jacobi-Bellman equation
associated to the optimal control problem. It can be linearized via
the Hopf–Cole transform, $\Phi\!=\!2D\,\log Z$,  by introducing the desirability function
$Z(\bm{x}_1,\theta_1;...;\bm{x}_N,\theta_N)$ \cite{todorov2009efficient}. Then 
the control \eqref{contrgeneral} becomes  
\begin{equation}
  f_i=2\,D\,\partial_{\theta_i}\ln Z\,,
\end{equation} 
that is a gradient ascent towards more desirable  configurations, hence the name.
Thanks to the Hopf–Cole transform, Eq.~\eqref{HJB} becomes the linearized Bellman equation
\begin{equation}\label{LB}
\frac{\lambda}{2\,D}\, Z-\frac{1}{4D}\sum_{i\neq j}\delta(\bm{x}_i-\bm{x}_j)\,(g_0-g_1\,\bm{n}(\theta_i)\cdot\bm{n}(\theta_j))\, Z+u_0\,\sum_{i}\bm{n}(\theta_i)\cdot\partial_{\bm{x}_i}\;Z+D\;\sum_i\partial^2_{\theta_i}\,Z=0\,.
\end{equation}
which is formally identical to the stationary Schr\"odinger equation
of $N$ identical, interacting bosons. We should solve both for the ground-state eigenvalue
$\lambda/2D$, which can be shown to be proportional to the total cost\footnote{ Note that, formally, the HBJ equation~\eqref{HJB} can be written as $C-\lambda-\mathcal{L}^\dagger \Phi=0$ with $\mathcal{L}^\dagger$ being the adjoint of the Fokker-Planck operator. Taking the average with respect to $P$ we get $\bar C-\lambda-\int P\mathcal{L}^\dagger \Phi=0$. Since, at the stationary point, $\mathcal{L}P=0$  holds, we can deduce that $\int P\mathcal{L}^\dagger \Phi=\int \Phi\mathcal{L} P=0$ from which $\bar C=\lambda$ follows.}, and the eigenfunction $Z$, requiring $Z$ to be real and
positive. To our knowledge, this quantum many-body problem has no known general solution for generic $N$; therefore, we will seek for the optimal control in an approximate mean field setting.

\subsection{Mean field approximation}\label{sec:MFA}
To simplify the problem and make it exactly solvable, we proceed with
a mean field approximation based on two hypothesis: first we assume
agent-wise factorization of the desirability $Z$ and then we assume
spatial homogeneity - no preferred points in space, only preferred
directions.  The agent-wise factorization
excludes direct pairwise interactions -- agents cannot directly dodge each other -- but, rather, each agent interacts
with the joint probability of the remaining $N-1$ ones in a self
consistent manner. This approximation 
is rather strong since in animal collective behavior it would make more sense to consider local interactions \cite{cavagna2018physics}. It must also be remarked that the homogeneity assumption
excludes from the description many interesting phenomena related to heterogeneities. Notwithstanding these limitations, we can still assume that this treatment could be relevant to
describe agents within a uniform bulk region of the swarm.

With the factorization and homogeneity assumptions, the desirability
can be written as
\begin{equation}
  \label{eq:desfact}
Z(\bm{x}_1,\theta_1,...,\bm{x}_N,\theta_N)=\prod_{i=1}^N\zeta(\theta_i)\,,
\end{equation}
and, equivalently
$\Phi(\bm{x}_1,\theta_1,...,\bm{x}_N,\theta_N)=\sum_{i=1}^N
\phi(\theta_i)$. As a consequence, the probability $P$ is factorized
as $P(\bm{x}_1,\theta_1,...,\bm{x}_N,\theta_N)=\prod_{i=1}^N
p(\bm{x}_i,\theta_i)$ and, owing to spatial homogeneity, we can write
$p(\bm{x}_i,\theta_i)=\frac{1}{V}\rho(\theta_i)$, with $V$ being the area
where the swarm moves.

We define the agents' average heading direction $\bar
\theta$ and the polar order parameter (alignment parameter or
polarization) $m$  as
\begin{equation}\label{m0}
m\,\bm{n}(\bar \theta)=\int d\theta' \; \bm{n}(\theta')\,\rho(\theta')\,,
\end{equation}
in terms of which the average agent speed reads $\langle \dot{ \bm{x}}\rangle=m\,u_0\, \bm{n} (\bar \theta)$; here and in the sequel, since all particles are equivalent by mean-field ansatz, we drop particle indices.

By defining the  parameter $\delta=(N-1)/V$, which is the
particle density measured by a reference agent, and $C_0=\delta\,
g_0/2$, with a few straightforward passages, we write the average
per agent cost as
\begin{equation}\label{barC}
\bar C=C_0+\int d\theta\;\rho(\theta)\;\left[-\frac{\delta\,m\,g_1}{2} \;\cos(\theta-\bar \theta)+\frac{1}{2}f^2\right]\,,
\end{equation}
and the functional~\eqref{functional} as
\begin{equation}\label{functionalMF}
\mathcal{H}=\lambda+\bar C-\int d\theta\;[\lambda+\phi(\theta)\;\mathcal{L}\,]\rho(\theta)\,,
\end{equation}
with $\mathcal{L}=-\partial_{\theta} f+D\,\partial_\theta^2$ being the single-particle Fokker-Planck operator as in Eq.~\eqref{FP0}. Proceeding analogously to the general case and by exploiting the Hopf-Cole transform with the factorized desirability \eqref{eq:desfact}, we derive the control to be
\begin{equation}\label{MFcontrol}
f(\theta)=2\,D\,\frac{d}{d \theta}\ln \zeta\,.
\end{equation}
Plugging the last expression into the stationary Fokker-Planck equation, $\mathcal{L}\,\rho=0$, with periodic boundary conditions, yields
\begin{equation}\label{norm}
\rho=\zeta^2\,,
\end{equation}
with $\int d\theta\;\zeta^2=1$. Therefore, the optimal control problem boils down to
solving the  self-consistent system of equations 
\begin{empheq}[left=\empheqlbrace]{align} 
&m=\int d\theta\; \cos\theta\;\zeta^2\label{m}\\
&\left[\frac{\lambda}{2D}+\frac{\delta \,m\,g_1}{2D} \;\cos\theta\right] \,\zeta+D\,\partial_\theta^2\zeta=0\,.\label{LBmf}
\end{empheq}
Equation~\eqref{m} is just Eq.~\eqref{m0} where we used \eqref{norm} and fixed $\bar \theta=0$ with no loss of generality, as the rotational symmetry can be broken in an arbitrary direction, while Eq.~\eqref{LBmf} is the mean-field linearized Bellman equation with $\bar \theta=0$. Formally, Eq.~\eqref{LBmf} is the stationary Schr\"odinger equation for a quantum pendulum also known as Mathieu equation that, for consistency with literature~\cite{brimacombe2020computation,gutierrez2003mathieu}, we rewrite in the canonical form
\begin{equation}\label{mathieu}
\left[a-2\,q \,\cos(2 \,y)\right] \,\zeta+\zeta''=0.
\end{equation}
with $y\!=\!\theta/2$, $a\!=\!-\!\lambda/(2\,D^2)$, $q\!=\!-\!\delta \,m\,g_1/D^2$,
and $''$ denoting the second derivative with respect to
$y$. Equation \eqref{mathieu} must be solved both for the
eigenvalue $-a$ (which corresponds to solving for $\lambda$) and the
eigenfunction $\zeta\!=\!\zeta_m$, which depends parametrically on
$m$. The solutions are the so-called Mathieu functions, as briefly recalled
in~\ref{app:Mathieu}. For periodic boundary conditions, the Mathieu
ground state eigenfunction, denoted as $\mathpzc{ce}_0$ in the literature, is
an even function with a single maximum in $y=0$ and a minimum in
$y=\pi/2$.  The associated eigenvalue is called characteristic Mathieu
function $a=a(q)$, which is  non-positive and takes the asymptotic expressions 
 $a(q)\sim -q^2/2$ (see Eq.~\eqref{a4}) and $a(q)\sim
2\,q+2\,\sqrt{-q}$ (see Eq.~\eqref{a_bigQ}) for  $q\to 0$ and $q\to -\infty$, respectively.

In order for an eigenfunction $\zeta_m$ to be a solution of the optimization problem, it must also satisfy the self-consistency condition~\eqref{m} which, using  Eq.~\eqref{LBmf}, reads
\begin{equation}\label{m1}
m=\int d\theta\,\cos\theta\,\zeta_m^2=\mathcal{F}(m).
\end{equation}
In the sequel we will drop the subscript in $\zeta_m$ whenever that
would not hinder clarity. 

\begin{table}[t!]
\centering
\begin{tabular}{ |p{3.5cm}|p{8cm}|  }
% \hline
% \multicolumn{2}{|c|}{Parameter table} \\
 \hline
Parameter& Description\\
 \hhline{|=|=|}
$D$   & rotational diffusivity\\
 \hline
 $g_0$ and $g_1$ & positional and angular collision cost\\
  \hline
 $\alpha$ & weight of the cost of control\\
 \hline
 $m$ & polar order parameter/polarization\\
 \hline
 $\delta=(N-1)/V$ & density of particles\\
  \hline
 $h=\delta\,g_1/D^2$ & tradeoff parameter\\
  \hline
 $q=-m\,h$ & Mathieu equation parameter\\ 
%   \hline
% $f$ & angular control\\ 
   \hline     
\end{tabular}
\caption{Summary of main parameters.}
\label{tab:parameters}
\end{table}

\begin{table}[t!]
\centering
\begin{tabular}{ |p{3.5cm}|p{8cm}|  }
 \hline
 \multicolumn{2}{|l|}{Mean field glossary} \\ \hline
 \hhline{|=|=|}
$\zeta$   & desirability\\
 \hline
 $\rho$ & single particle angular distribution \\
  \hline
 $\rho=\zeta^2$ & desirability \& angular distribution relation\\
 \hline
 $f=2D\frac{d}{d\theta}\ln \zeta$ & optimal control\\
 \hline
 $m=\mathcal{F}(m)$ & self-consistency equation\\

% $f$ & angular control\\ 
   \hline     
\end{tabular}
\caption{Summary of functions and main relations for the mean field model.}
\label{tab:functions}
\end{table}

\begin{figure*}[b!]
\centering
\includegraphics[width=.99\linewidth]{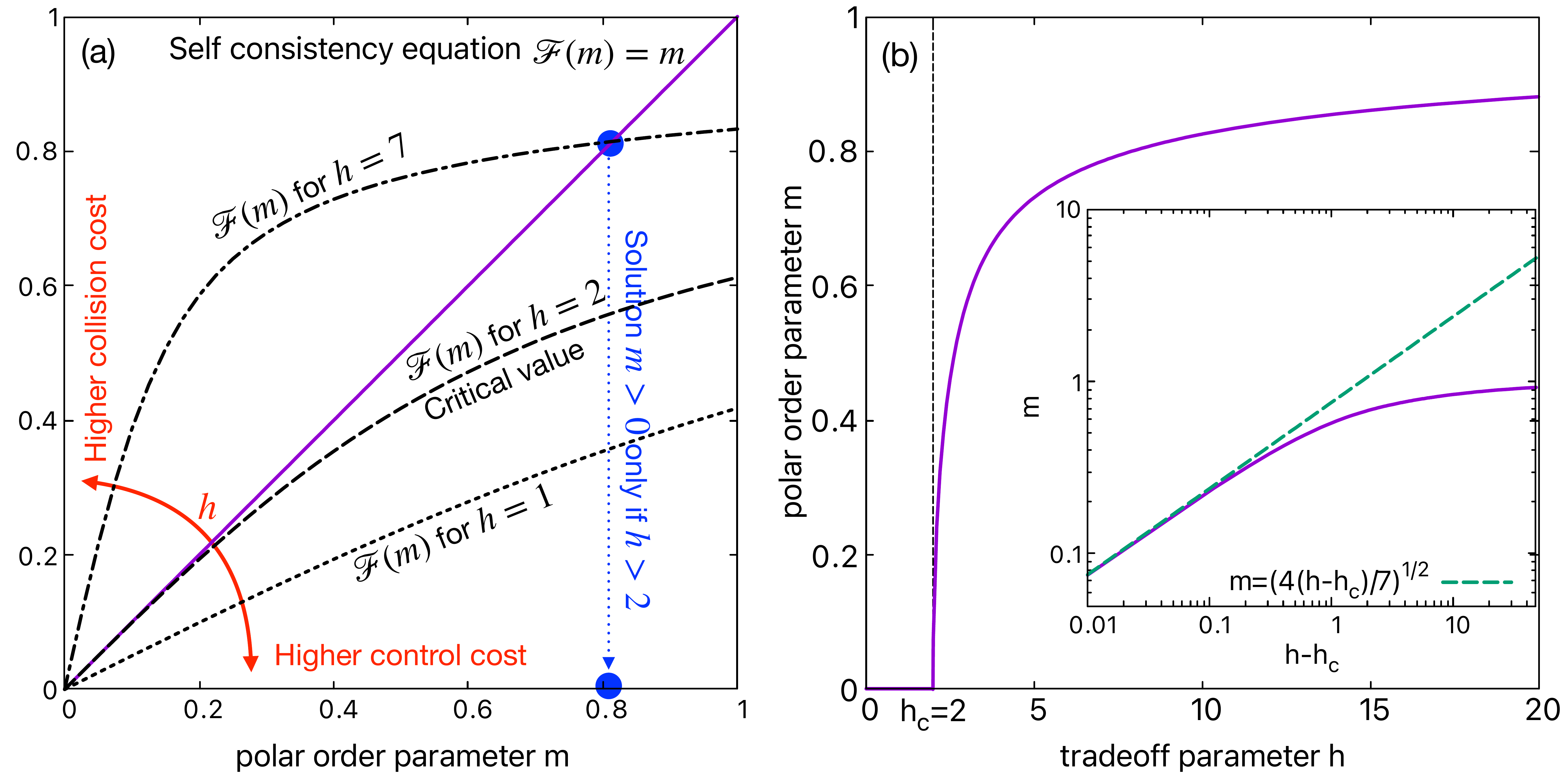}
\caption{\label{fig:uo} Self-consistency equation and polar order
  parameter: (a) graphical solution of Eq.~\eqref{m1},
  $\mathcal{F}(m)=m$ can only be satisfied with $m>0$ if
  $\mathcal{F}'>1$, which requires the tradeoff parameter
  $h=(N-1)\,g_1/(D^2\,V)$ to be larger than $h_c=2$; (b) polar order
  parameter $m$ as a function of $h$, a second order phase
  transition takes place at $h=h_c$. The inset displays the asymptotic
  approximation \eqref{crithopt} (green dashed line) showing the critical
  exponent to be $1/2$. }
\end{figure*}

The function $\mathcal{F}(m)$, shown in  Fig.~\ref{fig:uo}a, depends on the parameter (see Table~\ref{tab:parameters} for a handy summary of all the parameters of the problem and Table~\ref{tab:functions} for main relations and functions) 
\begin{equation}
h=-\frac{q}{m}=\frac{\delta\,g_1}{D^2}\,,
\end{equation}
which, besides containing all dependencies on the problem parameters
$N,\;g_1,\;D,\;V$, has a natural interpretation as the ratio between the parameters which control the importance of
collision and control costs. For a clearer picture, we  reintroduce
$\alpha$ and write as $h=(g_1\,\delta)/(\alpha \,D^2)$: at the
numerator, $g_1$ is rescaled with particle density $\delta$ while, at
the denominator, $\alpha$ is rescaled with diffusivity $D$. Note that
the solution does not depend on $g_0$, as follows from homogeneity
assumption.  As graphically illustrated in Fig.~\ref{fig:uo}a,
Eq.~\eqref{m1} admits only the trivial solution with no polar order
$m=m(h)=0$ for $0\leq h\leq h_c=2$, while a non-trivial solution
emerges, via a second order transition to non-zero alignment ($m>0$)
for $h> h_c$ (the critical value $h_c=2$ is derived in the next subsection by a perturbative expansion of the self-consistency equation).  The numerically computed function $m(h)$ is shown in
Fig.~\ref{fig:uo}b. The polar oder parameter $m$ is zero up to the
critical tradeoff value $h_c=2$, meaning that the alignment benefit
outweighs the cost of control only for $h>h_c$. Therefore, for
$h<h_c$, no control is applied and the system remains isotropic so
that the average cost $\bar C$ \eqref{barC} is equal to $C_0=\delta
g_0/2$ while, as derived in~\ref{app:costexpropt}, for $h>h_c$, $\bar
C$ is equal to
\begin{equation}\label{costexpropt}
\bar C=C_0+\frac{1}{2}D^2\;[h\,m^2(h)+a(-m(h)\,h)]\,.
\end{equation}
Note that, as anticipated, the constant $C_0$ is irrelevant to the
optimization process (as a consequence of the mean field assumptions),
while the remaining part depends only on the tradeoff parameter $h$,
up to the $D^2$ prefactor.

In the following sections, we will give a more detailed description of
both the critical behavior (for $h\to h_c=2$) and the strong coupling
(large $h$) regime.

\subsection{Critical behavior}\label{sec:CB}

As clear from Fig.~\ref{fig:uo}, at the critical point $h=h_c=2$,
there is a second order phase transition in the polar order parameter
with exponent 1/2, a classical mean field value, which can be
derived as follows. When $h=h_c^+$, we can solve Eqs.~\eqref{m}
and~\eqref{LBmf} perturbatively (see~\ref{app:OptiCrit}), by expanding
them in powers of $q=-m\,h$ for $q\to 0^-$ (small $m$ ansatz). Then,
we can write Eq.~\eqref{m} as
$m\,(h-2)/2-7\,(m\,h)^3/64+o(m^3)=0$, from which we obtain the asymptotic expression (for small $m$ or $\sqrt{h-h_c}$)
\begin{equation}\label{crithopt}
m=\sqrt{(4/7)\,(h-h_c)}\,,
\end{equation}
which, as shown in the inset of Fig.~\ref{fig:uo}b, fully captures the
critical behavior.  In this regime, the desirability $\zeta$ can be approximated at leading order in $m$  as
\begin{equation}\label{zcritic}
\zeta(\theta)=\frac{1}{\sqrt{2\,\pi}}\left[1+\frac{h\,m(h)}{2}\,\cos\theta\right]\,.
\end{equation}
Since the angular probability density function satisfies Eq.~\eqref{norm}, agents directions are uniformly distributed but for a tiny $O(m)$ cosine modulation. Plugging Eq.~\eqref{zcritic} into Eq.~\eqref{MFcontrol}, the optimal control at the critical point reads
\begin{equation}\label{controlCRIT}
f=-D\,h_c\,m\,\sin(\theta)+o(m)\,.
\end{equation}
Moreover, exploiting the asymptotic expressions \eqref{a4} for $a$ into  Eq.~\eqref{costexpropt},
we can obtain the cost close to the critical point $h-h_c= 0^+$:
\begin{equation}\label{asymptoptcost}
  \bar C-C_0=-(D^2/7)\,(h-h_c)^2+o((h-h_c)^2)\,.
\end{equation} 

We conclude the investigation of the critical properties by studying
the susceptivity to external perturbations, which is an important
observable in multi-agent systems, both when considering artificial
swarms control and biological collective behaviors, such as bird
flocks or insect swarms. Indeed, the susceptivity describes the crowd
sensitivity to fluctuations and/or external
stimuli~\cite{bialek2012statistical,cavagna2017dynamic}.  In order to define
the susceptivity, we need to specify an external field, and then compute
the derivative of $m$ with respect to it. In our setting, 
the external field is represented by a small collective nudge of
intensity $\epsilon$ in the direction $\hat \theta$, which formally
amounts to adding a small per-agent reward in the cost function
\begin{equation}\label{pert}
\delta G_i=-\epsilon\;\cos(\theta_i-\hat \theta)\,,
\end{equation}
for aligning along the direction $\hat\theta$. The perturbation \eqref{pert} breaks the isotropy favoring an average
alignment in the preferred direction $\hat \theta$ (which, with no loss of generality, can be set to
0). Since $\hat \theta$ breaks the isotropy, and the system has no intrinsic preferred direction, we can deduce that the system will polarize along the preferred direction $\bar \theta=\hat \theta$. Therefore, susceptivity is then defined as
\begin{equation}\label{chi}
\chi(h)=\left. \frac{\partial m}{\partial \epsilon}\right|_{\epsilon=0}\,.
\end{equation}
Essentially, the additional (negative) cost \eqref{pert} induces the shift
\begin{equation}\label{shiftepsilon}
q\mapsto q+\frac{2\epsilon}{D^2}\,,
\end{equation}
which, plugged into Eq.~\eqref{m1}, implicitly defines $m$ as a
function of $\epsilon$, for any $h$. Then, by a straightforward
application of Dini's implicit function theorem (see~\ref{app:sus}), we
obtain the following result: when $h<h_c$, $\chi=2/[D^2\,(h_c-h)]$
holds exactly; close to the critical point ($h\to
h_c^+$), $\chi=1/[D^2\,(h-h_c)]+o(1/(h-h_c))$. Therefore, near the critical
point, we have that the susceptivity diverges with $h \to h_c$ as
\begin{equation}
\chi\sim |h-h_c|^{-1}\,,
\end{equation}
we notice that the critical exponent $1$ for the susceptivity is also quite standard in mean field theories.
On the other hand, exactly at the critical point $h=h_c$, $m$ depends on $\epsilon$  as $m\sim 2\,\left[\frac{\epsilon}{7\,D^2}\right]^{1/3}$ and, hence, $m$ is a continuous but non differentiable function of $\epsilon$ at the critical point and, consequently, the susceptivity diverges as
 \begin{equation}
 \chi\sim \epsilon^{-2/3}\,.
 \end{equation}
The absence of a first order discontinuity implies that, at the critical point, an infinitesimal nudge
is not enough to induce finite polarization in this kind of system. This a natural consequence of the continuous symmetry which is spontaneously broken even in absence of external perturbations. On the other hand, any small nudge is enough to fix the direction of the polarization vector.

\subsection{Strong coupling}\label{sec:SC}
The strong coupling regime is identified by the condition $h\gg h_c$,
which can be achieved with high density ($\delta=(N-1)/V\gg1)$, with a
high collision cost coefficient $g_1$ or with low noise $D$. In the large $h$ limit,
we can analytically  solve Eqs.~\eqref{m} and~\eqref{LBmf} 
by exploiting a small oscillations ansatz which transforms 
 the quantum pendulum into a quantum harmonic oscillator
 \eqref{MathieuOscill}. As discussed in \ref{app:optlargeh},  in this limit $m$
 approaches to 1 asymptotically (strong alignment) as
\begin{equation}\label{eq:mstrong}
 m\sim
 1-1/2\,\sqrt{h}\,.
\end{equation}
 The solution $\zeta$ becomes
\begin{equation}
\zeta(\theta)\sim \left(\frac{2\sqrt{h\,m(h)}}{\pi}\right)^{1/4}\,\exp(-\sqrt{h\,m(h)}\,\theta^2)\,,
\end{equation}
which via Eq.~\eqref{norm} implies that most agents are aligned along the average direction $\bar \theta$ and that - consistently with the small oscillations assumption - normal fluctuations vanish as $h^{-1/4}$. The associated asymptotic control is linear in $\theta$
\begin{equation}\label{controlLARGE}
f=-D\,\sqrt{h\,m(h)}\;\theta\,,
\end{equation}
where $m\approx 1$ from \eqref{eq:mstrong}.  Note that, since this
solution has been obtained in the small oscillations regime and,
therefore, Eq.~\eqref{controlLARGE} is only accurate around $\theta=0$
which is, on the other hand, the only region which matters, since the
probability of visiting other regions is exponentially suppressed.
Exploiting the large $h$ asymptotics for the Mathieu characteristic
function \eqref{a_bigQ}, one can easily obtain that the cost decreases
approximately linearly with $h$ as $\bar C-C_0=
D^2\,[-h/2+(3/4)\,\sqrt{h}]+o(\sqrt{h})$.  Using the same scheme of
the previous section, we can compute the susceptivity is this regime
as well obtaining that $\chi$ \eqref{chi} vanishes for large $h$ as
$\chi\sim 1/[2\,D^2\,h^{3/2}\,m^{1/2}]$ (see~\ref{app:sus}).
Therefore, as we might have expected, the polar order strength $m$ is
little responsive to external nudges, when it is already close to its
maximum value 1.

\section{Sinusoidal control vs optimal solution}

\subsection{Motivation}\label{sec:motsin}

The optimal solution \eqref{controlCRIT} to the collision avoidance
problem in swarming active Brownian particles suggests that close to
the critical point the optimal control is sinusoidal at leading order,
i.e. 
\begin{equation}\label{controlSin}
f=-D\,K\;\sin(\theta-\bar\theta)\,,
\end{equation}
with $K\approx h_c m$; notice that in Eq.~\eqref{controlCRIT} the mean
field direction was put to zero ($\bar\theta=0$) for the sake
simplicity and here restored for clarity. Interestingly, also the
strong coupling optimal solution \eqref{controlLARGE} is well
approximated by the sinusoidal control \eqref{controlSin}: owing to
the small oscillations property $|\theta-\bar\theta|\ll1$, we can
replace $\theta-\bar\theta$ with $\sin(\theta-\bar\theta)$ in
Eq.~\eqref{controlLARGE}, obtaining $K\approx \sqrt{hm}$.  Therefore,
in both asymptotics ($h\to h_c^+$ and $h\to\infty$),
Eq.~\eqref{controlSin} approximates the optimal control with a
rescaled control strength $K$, which only depends on the parameter $h$ in both
cases. This is quite noteworthy as the search of the optimal control
is done without specifying the functional form of the control.

Remarkably, the control \eqref{controlSin} is reminiscent of the mean
field versions of the Kuramoto model \cite{kuramoto1975self} with zero
natural frequencies, which is a paradigm for synchronization
\cite{acebron2005kuramoto}, and of the (time-continuous) stochastic
Vicsek model 
\cite{peruani2008mean,chepizhko2009kinetic,chepizhko2010relation,chepizhko2021revisiting}, which is a variant
of one of the most popular models used for describing collective
motions and swarming of self-propelled agents
\cite{vicsek2012collective,ginelli2016physics}. In the mean field
version of the latter \cite{peruani2008mean,chepizhko2009kinetic,chepizhko2021revisiting}, individual agents
are driven by the approximate control $f_{V}=-R\,m\,\sin(\theta-\bar
\theta)$, corresponding to Eq.~\eqref{controlSin} upon defining
\begin{equation}\label{Rcontro}
R=D\,K/m
\end{equation}
whenever the polar order parameter $m=m(K,D,g_1,V)$ is non zero.

Given the similarity, both in the critical and in the strong coupling
limit, of the optimal control with the classical models discussed
above it is worth to compare \textit{vis a vis} the optimal control
solution with the class of ``sinusoidal control models'' defined by
Eq.~\eqref{controlSin}. In particular, we aim at comparing the optimal
control with the ``best'' sinusoidal control, defined by
Eq.~\eqref{controlSin} with $K=K_\star$, with $K_\star$ being the
value of $K$ that minimizes the average cost, given the parameters of
the problem $(N,g_1,V,D)$. As we will see, actually best sinusoidal
model will depend on the familiar combination $h=(N-1)g_1/(VD^2)$, so that
$K_\star=K_\star(h)$ is a function of the tradeoff parameter
$h$ only.

In the following subsection, after obtaining the best sinusoidal
control, we briefly discuss the critical and strong coupling regimes
and, then, we compare the optimal and best sinusoidal control for arbitrary tradeoff parameter values.

\subsection{Best sinusoidal model}\label{sec:BSM}
\begin{figure*}[h!]
\centering
\includegraphics[width=1\linewidth]{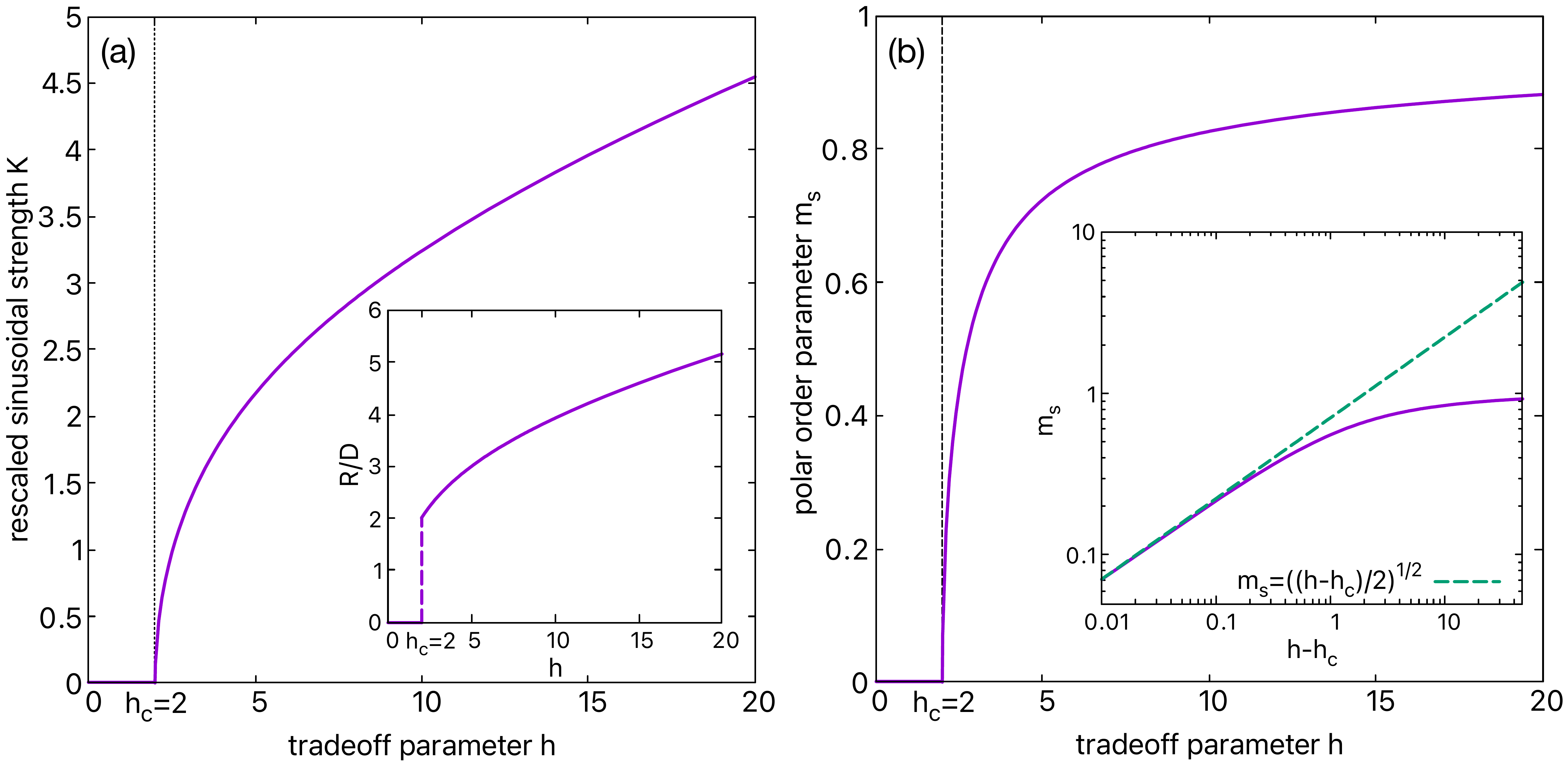}
\caption{\label{fig:uv} (a) Rescaled best sinusoidal control strength
  $K_\star$ as a function of the tradeoff parameter $h$. It is
  non-zero only for $h>h_c=2$, where a second order transition takes
  place. The inset shows the control strength
  $R_\star(h)=K_\star(h)/(m_sD)$ of the associated Vicsek mean field
  model, see Eq.~\eqref{Rcontro}. (b) polar order parameter $m_s$ as a
  function of $h$ for the best sinusoidal control. A second order
  phase transition is visible at $h=h_c=2$ with critical exponent
  $1/2$. The inset shows the asymptotic approximations \eqref{eq:ms}
  (green dashed line) valid close to the critical point. Notice
  the similarity with Fig.~\ref{fig:uo}b. }
\end{figure*}

The dynamics of the heading direction for a generic agent with the
sinusoidal control \eqref{controlSin} is $d\theta=-K\,D\, \sin\theta dt+
\sqrt{2D} \,d\xi$, where $\xi$ is the usual Wiener noise and where,
again, we assume $\bar\theta=0$ for the sake of notation
simplicity. Note that the above dynamics also describes the
orientation of gravitactic (bottom-heavy) microorganisms in two
dimensions, where $1/KD$ is the time scale with which the organism
orients vertically upward oppositely to gravity
\cite{pedley1987orientation,cencini2016centripetal}.  The
Fokker-Planck equation associated with such dynamics,
$\partial_\theta(-KD\sin\theta-D\partial_\theta)\rho_s=0$ is solved
by the Fisher-Von Mises distribution~\cite{mardia2009directional}
\begin{equation}\label{VonMises}
\rho_s(\theta)=\frac{1}{2\,\pi\,I_0(K)}e^{K\,\cos\theta}\,,
\end{equation}
where $I_\alpha(z)$ is the modified Bessel function of the first kind of
order $\alpha$ (briefly surveyed in \ref{app:Bessel}) and where the
subscript $s$ is used to remind that it pertains to the sinusoidal
control. Once the distribution is known, we can compute the polar order parameter as
\begin{equation}\label{mK}
m_s(K)=\int d\theta \rho_s(\theta) \cos\theta=\frac{I_1(K)}{I_0(K)}\,.
\end{equation}
As stated, the best sinusoidal model is given by Eq.~\eqref{controlSin} with such $K_\star$ that minimizes the average cost \eqref{barC}. By plugging Eqs.~\eqref{VonMises} and ~\eqref{controlSin} into Eq.~\eqref{barC}, after a few trigonometric passages combined with the identity~\eqref{bess2}, the sinusoidal control cost $\bar C_s$ can be written as 
\begin{equation}\label{costSin}
\bar C_s[K,D]=\int d\theta\,\rho_s\,C_s=C_0+D^2\left[-\frac{h}{2}\,m_s^2(K)+\frac{K}{2}\,m_s(K)\right]\,,
\end{equation}
from which one can easily deduce that $K_\star=K_\star(h)$, as anticipated.
Unfortunately, it is impossible to minimize the cost~\eqref{costSin} analytically with respect to $K$, so we proceeded numerically. In Fig.~\ref{fig:uv}a we show both $K_\star(h)$  as a function of the tradeoff parameter $h$. 

Figure~\ref{fig:uv}b displays the behavior of the polar order
parameter $m_s=m_s(h)$ as a function of the tradeoff parameter $h$.
Like in optimal case, the polar order parameter displays a second
order phase transition at the same critical point $h=h_c=2$ and with the same
critical exponent $1/2$.  Indeed, as
derived in \ref{app:SinCrit}, by expanding the cost $\bar C_s$ for
small $K$ one can derive that for $h-h_c\ll 1$, at leading order
\begin{equation}
  K_\star\approx \sqrt{2(h-h_c)}\,,
\end{equation}  
and that (see also inset of Fig.~\ref{fig:uv}b)
\begin{equation}
m_s(h)\approx \sqrt{(1/2)(h-h_c)}\,,
\label{eq:ms}
\end{equation}
so that the sinusoidal control~\eqref{controlSin}, near the critical
point, features a rescaled strength $K_\star\approx h_c m_s$ which is
similar to the optimal one \eqref{controlCRIT}. Note, however, that
$m$ and $m_s$ are not fully equivalent. The previous expressions also
imply that the best coupling $R_\star$ (Eq.~\eqref{Rcontro}) from
the Vicsek interpretation displays a first order discontinuity, as
shown in the inset of Fig.~\ref{fig:uv}a.

In spite of the similarities,  the
asymptotic dependence of $m_s$ on $h$ (for $h\to h_c^+$) differs by a
pre-factor with respect to the optimal control: indeed comparing
Eqs.~\eqref{eq:ms} and \eqref{crithopt} we have different prefactors
$\sqrt{1/2}$ and $\sqrt{4/7}$, respectively which differ by a mere
$6\%$.  At a first glance this difference may seem surprising, but
 it can actually be rationalized by observing that the sinusoidal
control is just a first order approximation to the optimal one while, as detailed below, the polar order parameter prefactor at criticality is
determined  by the first sub-leading order.

For the optimal control, we have derived the optimal critical behavior by
expanding the self-consistency condition \eqref{m1}
$\mathcal{F}(m)=m$. The latter can be rewritten as
$m=\langle\cos\theta\rangle_m $, which plays the same role as
Eq.~\eqref{mK} for the sinusoidal model; $\langle [\ldots]\rangle_m$
is to remind that the probability density depends on the polarization
itself as typical in self-consistent problems. As the control is odd
w.r.t. the transformation $\bar \theta\mapsto \bar \theta+\pi$, it
turns out that $\langle\cos\theta\rangle_m$ is odd in
$m$. Consequently, close to the critical point (i.e. form small $m$)
we can write
$\langle\cos\theta\rangle_m=c_1\,hm+c_k\,(h\,m)^k+o((hm)^k)$ with $k>1$
being an odd integer, which we know to be 3 (see
e.g. Eq.~\eqref{eqforubar_order3}). Now, imposing the self-consistency
condition $m=\langle\cos\theta\rangle_m$ yields
\begin{equation}
m\approx\left[\frac{1}{h^k\,c_k}\,(1-c_1\,h)\right]^{\frac{1}{k-1}}\,,
\end{equation}
from which we deduce that the critical point $h_c=c_1^{-1}$ is solely
determined by the first order and is, therefore, a leading order
effect; the critical exponent is fixed by the value of $k$ (ordinal number of the first non-vanishing sub-leading order)
as $1/(k-1)$ ($1/2$ in our case as $k=3$) and the prefactor 
is given by the sub-leading order prefactor $-c_1/(c_k\,h^3)$. Since the optimal
solution is sinusoidal at leading order, we expect $h_c=2$ to hold for both
the optimal and best sinusoidal controls, by construction. The value of
$k=3$ and, therefore, the critical exponent $1/2$, is fixed by the
symmetry and should be the same for both models. However, any further
equivalence is not obvious and, in particular, there is no specific
reason for $c_3$ to be the same: they are actually different and this
explains the difference in the prefactors discussed above.

Now we briefly discuss the strong coupling regime. For large $h$
(which implies large $K$), the Von-Mises distribution
\eqref{VonMises} is well approximated by the Gaussian
\begin{equation}
\rho_K=\sqrt{\frac{K}{2\pi}}\;\exp\left(-\frac{K}{2}\theta^2\right)\,,
\end{equation}
and $\cos\theta\approx 1-\theta^2/2$. With such an approximation the
self-consistency condition~\eqref{mK} yields
$m_s=1-\frac{1}{2\,K}+o(1/K)$, which inserted into the cost
\eqref{costSin} and minimizing with respect to $K$ gives
\begin{equation}
 K_\star= \sqrt{h}+o(\sqrt{h})\,.
\end{equation}
Consequently, the polar order parameter in the large $h$ limit is given by
$m_s\approx 1-1/2\sqrt{h}$ as for optimal case \eqref{eq:mstrong}.

The above discussion establishes a connection between sinusoidal
and optimal model in the asymptotic regimes, but provides little insights
into the intermediate region. In the next section, we further
investigate the differences and similarities between optimal and best
sinusoidal control.

\subsection{Comparison between optimal solution and best sinusoidal model}\label{sec:CBOSABSM}

\begin{figure*}[b!]
\centering
\includegraphics[width=1\linewidth]{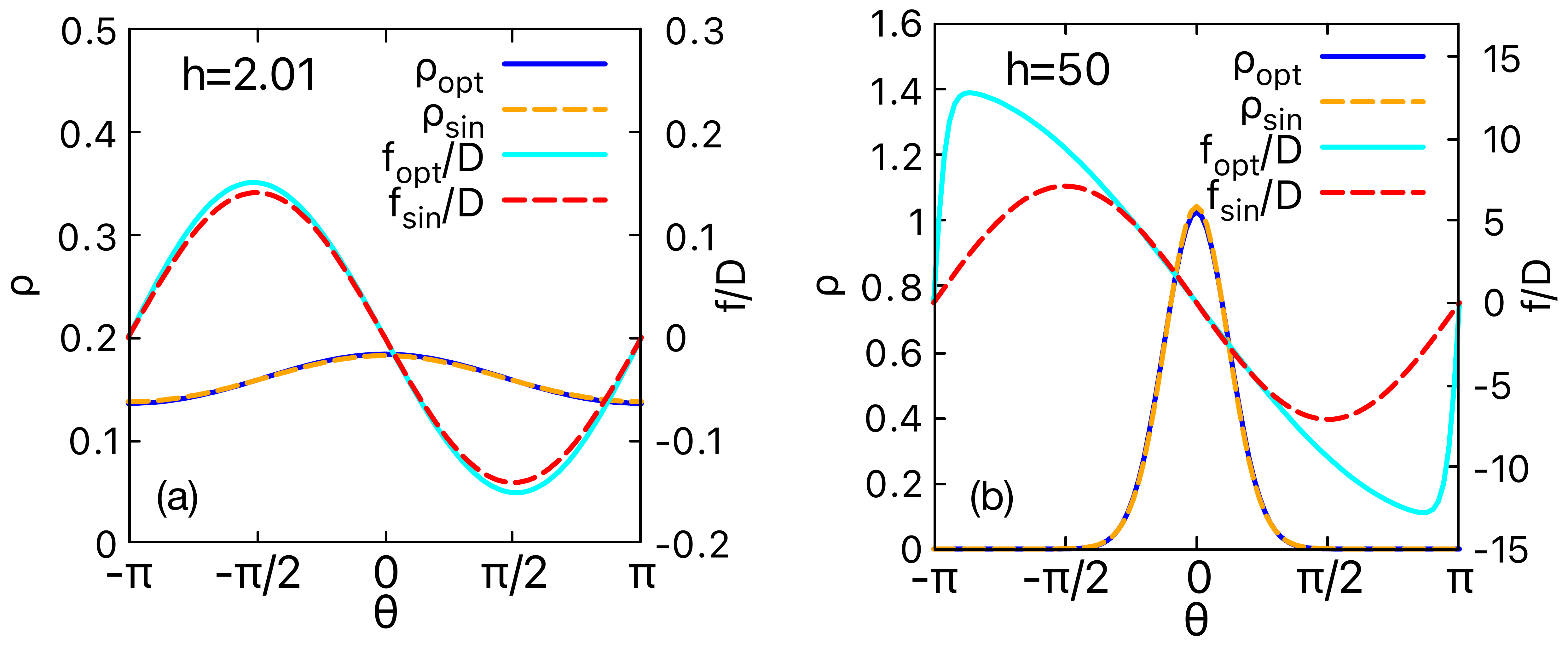}
 \caption{\label{fig:h2d1_control_and_pdf} Comparison for the optimal
   and best sinusoidal model between the angular probability density
   function (left scale) and the control $f/D$ (right scale) for (a)
   $h=2.1$, i.e. close to the critical point, and (b) $h=50$,
   corresponding to the strong coupling regime. See the figure legend
   for the various curves.}
\end{figure*}

\paragraph{Critical case comparison}

\begin{figure*}[t!]
\includegraphics[width=0.99\linewidth]{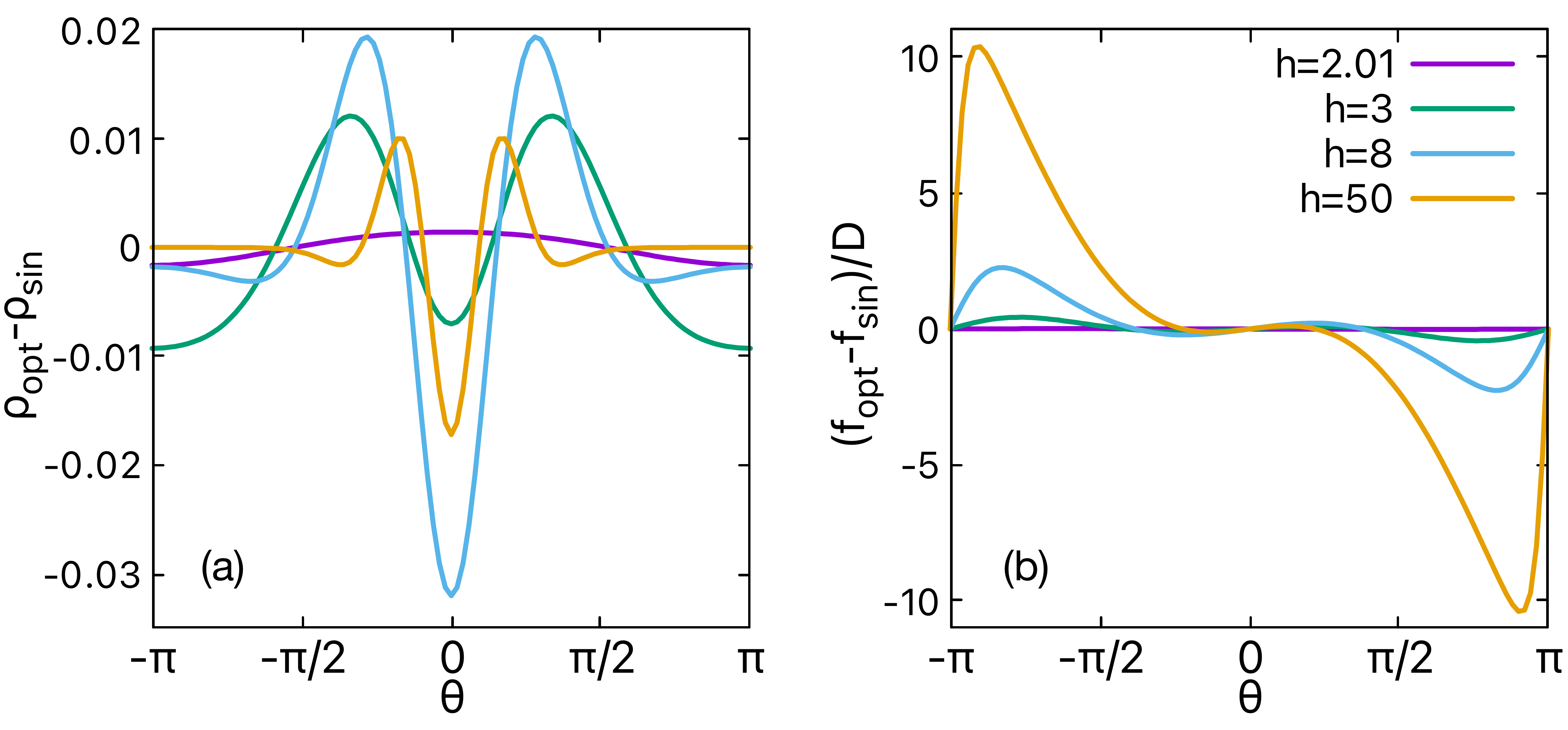}
 \caption{\label{fig:pdfdiff} Differences between the optimal and best
   sinusoidal model: (a) angular probability density and (b) rescaled
   control, $(f(h)-D\,K_\star(h)\,\sin(\theta))/D$, for different $h$
   as in figure legend.  }
\end{figure*}

\begin{figure*}[b!]
\centering
\includegraphics[width=1\textwidth]{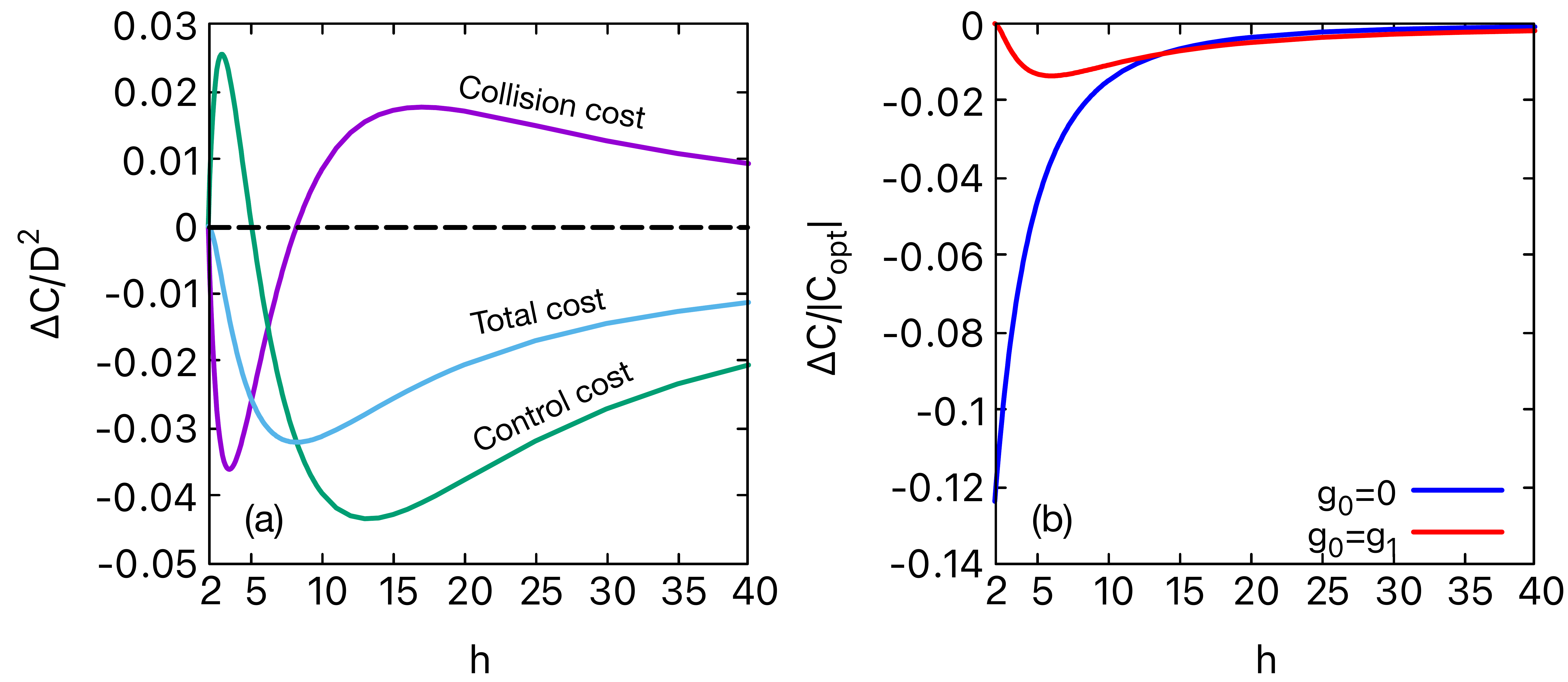}
\caption{\label{fig:costs_compared} Cost comparison between optimal and best sinusoidal model: (a) rescaled average cost differences $\Delta \bar C=(\bar C-\bar C_s)/D^2$ - total (cyan curve), collision (purple) and control (green)- plotted as universal functions of the tradeoff parameter $h$; (a) the relative cost difference $\Delta \bar C/|\bar C|$,  as a function of the tradeoff parameter $h$ in two notable cases $g_0=0$ (pure alignment) and $g_0=g_1$ (pure collision). 
 }
\end{figure*}

For both models, below the critical point, $h<h_c$, no control is
exerted as it is too expensive. Consequently, the stationary distribution of agents
orientation is uniform in both cases. Near the critical point,
$h-h_c=0^+$, the angular distribution remains approximately uniform
and collision costs remain high, since collisions of anti-aligned
particles are common. However, in this regime, tiny deviations of the
optimal control from the sinusoidal one (see
Fig.~\ref{fig:h2d1_control_and_pdf}a) allows to slightly squeeze the
distribution towards the alignment with the mean direction $\bar
\theta=0$ in Fig.~\ref{fig:h2d1_control_and_pdf}b and
\ref{fig:pdfdiff}a. Therefore, the optimal model pays slightly more in
the control cost to achieve a reduction of the collision cost
w.r.t. the sinusoidal one, overall reducing the average cost, as shown
in Fig.~\ref{fig:costs_compared}.  In particular, close to the
critical point, the cost difference between the optimal and sinusoidal
model is
\begin{equation}\label{costdiffcrit}
\Delta \bar C=\bar C-\bar C_s=-\frac{D^2}{56}\,(h-h_c)^2+o((h-h_c)^2)\,,
\end{equation}
as obtained by subtracting Eq.~\eqref{csin} from  Eq.~\eqref{asymptoptcost}.

\paragraph{Intermediate and strongly interacting regimes}
As $h$ grows away from the critical point, both distributions shrink
towards the origin, but not exactly in the same way. The best
sinusoidal model distribution is more peaked both around the origin
(strong alignment) and around $\pm \pi$ (strong anti-alignment), while
intermediate values are less probable. This difference, highlighted in
Fig.~\ref{fig:pdfdiff}a, is the largest around $h=\hat{h} \approx 8$,
which also corresponds to the the region where the difference between
the total costs $\Delta \bar C$ is the largest, as shown in
Fig.~\ref{fig:costs_compared}a. Just before $\hat{h}$, the optimal
model outperforms the sinusoidal one in both control and collision
costs. However, the collision costs advantage rapidly declines and,
for large $h$, the edge of the optimal solution is preferred only due
to lower control costs, while collision costs are higher. To
understand the origin of these differences, we should look at the
shape of the optimal control, which starts sinusoidal at $h-h_c=0^+$
and then approaches a sawtooth shape for $h\gg 1$ (see
Fig.~\ref{fig:pdfdiff}b and \ref{fig:h2d1_control_and_pdf}b). The strong
control near $\pm\pi$ makes little difference in terms of costs,
because the probability of exploring such region is exponentially
suppressed.  Indeed, both optimal and best sinusoidal distributions
converge to the same Gaussian distribution with vanishing variance for
large $h$ (see Fig.~\ref{fig:h2d1_control_and_pdf}b). In other terms, the two
seemingly different controls (Fig.~\ref{fig:pdfdiff}b) only contribute
with their linear approximation near the origin and are therefore
equivalent, consistently with derivation of the previous section.

\paragraph{Rescaled-cost difference analysis }
We have provided a detailed description of the differences between the
two models. We should remark that all discrepancies we have
highlighted are somehow small, since the two distributions never show
significantly different shapes. Moreover, at the critical point, the
critical exponents are the same and, finally, the polar order parameters are
closely related for all $h$. The most delicate point is the cost
difference: the universal behavior of the average cost difference as
a function of $h$ (Fig.~\ref{fig:costs_compared}a) only emerges when
rescaling with such difference with the factor $D^2$. In other
terms, such difference can be made arbitrarily large as can be
deduced, for instance, by rescaling $g\mapsto z\,g$ and $D\mapsto
D\,\sqrt{z}$. Under this transformation, while $h$ does not change, the
cost difference does, as $\Delta \bar C\mapsto z\,\Delta\bar C$.

\paragraph{Relative-cost difference analysis }
The above observation implies that a sound analysis should take into
account the relative difference $\Delta \bar C/\bar C$. Note, though,
that such quantity depends on the constant $C_0$ and thus on $g_0$
(which otherwise play no role in the optimization process). However,
as discussed $g_0$ relates to the interpretation of the model, thus it
is not possible to give a universal description. We can briefly
consider some notable cases: $g_0=0$ and $g_0=g_1$ which represent a
pure alignment and collision problem, respectively. In the former
case, $\Delta \bar C/|\bar C|$ is 0 for $h<h_c$ and then (first order
discontinuity) jumps to $-1/8$ at $h=h_c$ (see
Fig.~\ref{fig:costs_compared}b) and finally vanishes to zero for
$h\to\infty$. Note, however, that the maximal relative cost difference
at $h=h_c^+$ is obtained in a limit in which both $\Delta \bar C$ and,
mostly important, $\bar C$ itself vanishes (see
Eqs.~\eqref{costdiffcrit} and~\eqref{asymptoptcost} with $C_0=0$). On
the other hand, in the $g_0=g_1$ case, we have that $\bar C=C_0\neq 0$
at $h=h_c$ meaning that the even the cost of isotropic random
collisions is not zero. Consequently, since the denominator never
vanishes, we have $\Delta \bar C/|\bar C|\to 0$ for $h\to
h_c^+$. Similarly, for $h\to\infty$, it is easy to see that $\Delta
\bar C/|\bar C|\to 0$, as the two models are equivalent at leading
order. Therefore, the maximal relative difference is realized for some
$h\in(h_c,\infty)$: a numerical test shown in
Fig.~\ref{fig:costs_compared}b reveals that this is realized at
$h\approx 6$ with $\Delta \bar C/|\bar C| \approx -0.0135$. By
considering both the rescaled and the relative cost difference
analysis, we can conclude that the sinusoidal model is always a good
approximations for the optimal solution in realistic scenarios.

\section{Discussion\label{sec:end}}

We have shown by means of optimal control techniques, that the optimal
behavior for collision-avoiding active particles can be characterized
by a tradeoff parameter $h$ between collision and control costs, and
that the polarization of the system undergoes a phase transition in
the mean-field regime. The possibility of approaching this problem
analytically, albeit in an approximate form, provided insights into
the features of the optimal solution and a comprehensive statistical
characterization. Moreover, we found that the optimal behavior, both
close to the transition and for large tradeoff values, is well
approximated by a mean-field version of the kinetic Vicsek model
\cite{peruani2008mean,chepizhko2009kinetic,chepizhko2021revisiting}
which also displays a second order transition. Therefore, such model,
whose short range version was mainly derived from phenomenology, turns
out to be a quasi-optimal solution for the collision-avoidance task,
given appropriate parameters. Clearly, when working with task-oriented
agents, as in biological systems or robotics, being close to
optimality is a highly desirable feature. The optimal control
framework is therefore a very valuable tool not just for discovering
new optimal models, but also for assessing the quality of existing
ones with respect to some performance criteria. Moreover, in
accordance with previous works,
e.g. Ref. \cite{pezzotta2018chemotaxis} where known chemotactic
behaviors were found to be optimal solutions to target search problem,
our findings suggest that optimal control formalism can, at least in
some cases, provide a theoretical ground to interpret some biological
solutions in situations where specific tasks need to be solved.

While our analysis was restricted to a single scenario, the same
approach could be successfully carried in different settings to
explore other classes of collective behaviors, for instance allowing
for linear acceleration or for particles of finite sizes. Also,
remaining in the context of mean-field Vicsek-like models, it would be
interesting to explore how different kinds of noise can influence the
optimal solution.  Another interesting outlook would be to go beyond
the mean field approximation, either analytically or numerically, by
introducing a spatial structure.

\section*{Acknowledgments}
We thank A. Cavagna and S. Pigolotti for useful comments on the manuscript.

\appendix
\section*{Appendices}

\section{Mathieu functions }\label{app:Mathieu}

Mathieu functions are solutions of eigenvalue Schr\"odinger equation
for the quantum pendulum, which is customarily written as
\begin{equation}\label{eqmathieuapp}
\zeta''+(a-2\,q\,\cos(2\,y))\,\zeta=0\,,
\end{equation}
with $y\in[-\pi/2,\pi/2]$, and where $q$ is some constant and $a$ is
(minus) the eigenvalue (the ``energy''). For any $q$, the eigenvalues
$a_k(q)$ depend on a parameter $k$ and are called Mathieu
characteristic function. By imposing boundary conditions, $k>0$
becomes and integer representing the ordinal number of the energy
level: if $q<0$ then $-a_k(q)<-a_{k+1}(q)$. The eigenfunctions
$\mathpzc{m}_k$ are either even $\mathpzc{ce}_{k}=\mathpzc{m}_{2k}$ or
odd $\mathpzc{se}_k=\mathpzc{m}_{2k+1}$. Since we are only interested
in the ground state with periodic boundary conditions, we focus on
$k=0$. The eigenvalue is $a(q):=a_0(q)$ (main text notation) and the
eigenfunction $\zeta=\mathpzc{ce}_{0}(a(q),q)$. For further details,
see for instance
Refs.~\cite{brimacombe2020computation,gutierrez2003mathieu,gradshteyn2014table}.

\section{Optimal solution near the critical point}\label{app:OptiCrit}
In this appendix, we derive the mean-field solution of the system of
equations \eqref{m} and \eqref{LBmf} near the critical point.
We start from the Mathieu equation \eqref{eqmathieuapp} and we observe
that $m\to 0$ implies $q=-h\,m\to 0$ (see Table~\ref{tab:parameters})
for finite $h$. We solve the equation perturbatively by expanding the
solution $\zeta$ and the eigenvalue $a$ in power series of $q$, by writing
\begin{equation}
\begin{dcases}
\zeta=\zeta_0+\zeta_1+\zeta_2+\zeta_3+\zeta_4+o(q^4)\\
a=a_0+a_1+a_2+a_3+a_4+o(q^4).
\end{dcases}
\end{equation}
where $a_k=O(q^k)$ and $\zeta_k=O(q^k)$ (for instance in $C_\infty$ norm). At any order $k$, we impose $\zeta''+(a-2\,q\,\cos(2\,y))\,\zeta=o(q^k)$ and $\int_{-\pi/2}^{\pi/2} dy\, \zeta^2=1+o(q^k)$, starting from order 0 to 4. Clearly, for any $k$, $\zeta_k$ and $a_k$ only depend on $\{\zeta_s\,,a_s\}_{s<k}$. From this procedure, we find that the eigenvalue is
\begin{equation}\label{a4}
a(q)=-\frac{1}{2}\,q^2+\frac{7}{128}\,q^4+o(q^4),
\end{equation}
while the eigenfunction can be written as the following expansion in harmonics
\begin{equation}\label{fullzetaexpr3}
\zeta=\frac{1}{\sqrt{\pi}}\left[1-\frac{q^2}{16}+\left(-\frac{q}{2}+\frac{11\,q^3}{128}\right)\,\cos(2\,y)+\frac{q^2}{32}\,\cos(4\,y)-\frac{q^3}{1152}\,\cos(6\,y)\right]+o(q^3).
\end{equation}
 From expansion~\eqref{fullzetaexpr3}, the self-consistency condition
 \eqref{m1} can be written as
\begin{equation}\label{eqforubar_order3}
m=\int_{-\pi/2}^{\pi/2}dy\,\zeta_0^2\,\cos(2 \,y)=\frac{h\,m}{2}-\frac{7\,( m\,h)^3}{64}+o(m^3)\,,
\end{equation}
which has 3 solutions: the trivial one $m=0$, an unphysical solution $m<0$ and
\begin{equation}\label{m_vs_h_opt}
m\sim\sqrt{\frac{32}{7\,h^3}}\sqrt{h-2}\sim \sqrt{\frac{4}{7}}\sqrt{h-2}
\end{equation}
which is only well defined for $h>h_c=2$, where a second order phase transition occurs. This yields the asymptotic behavior and the critical exponent $1/2$. 

\section{Optimal solution for large $h$.}\label{app:optlargeh}
To study the $h\gg h_c$ regime, we assume $m\approx 1$, from which it
follows that $q(h)=-h\, m(h)$ becomes large negative. For large $q$, then both
$\zeta$ and $\rho$ are peaked around $y=0$ and, therefore,
we can assume a regime of small oscillations $|y|\ll
1$.  Then, by
expanding $\cos(2\,y)\approx 1-2\,y^2$, as one would expect from physics, Eq.~\eqref{eqmathieuapp}) reduces to the well-known Schr\"odinger equation of the 
quantum harmonic oscillator: 
\begin{equation}\label{MathieuOscill}
-\frac{1}{2}\zeta''+\frac{1}{2}\underbrace{(-4\,q)}_{\omega_0^2}\,y^2\,\zeta=\underbrace{\frac{a-2\,q}{2}}_{E}\,\zeta.
\end{equation}
From the ground state eigenvalue solution $E_0=\omega_0/2$, we get the large $q$ approximation of the characteristic function $a$
\begin{equation}\label{a_bigQ}
a(q)\approx 2\,q+2\,\sqrt{-q}\mbox{ for }q\ll-1\,.
\end{equation}
Conversely, from the ground state eigenfunction $\exp(-\omega_0/2\,y^2)$, we get
\begin{equation}
\zeta=\left(\frac{2\sqrt{-q}}{\pi}\right)^{1/4}\; \exp(-\sqrt{-q}\,y^2)
\end{equation}
where we have extended the domain of $y$ from $[-\pi/2,\pi/2]$ to
$(-\infty,\infty)$ by enforcing normalization $\int_{-\infty}^{\infty}
dy\;\zeta^2(y)=1$. We can then rewrite Eq.~\eqref{m1} as
$\int_{-\infty}^{\infty} dy\;\zeta^2(y)\,\cos(2\,y)=m$ which becomes
\begin{equation}\label{u_and_h_gauss}
m=e^{-\frac{1}{2 \sqrt{h m}}}.
\end{equation}
Hence, if $h\to\infty$, then $m\to 1$, validating our small-oscillations ansatz. More precisely
\begin{equation}\label{u_vs_h}
1-m\sim \frac{1}{2\,\sqrt{h}}\mbox{ for }h\to\infty\,.
\end{equation}

\section{Proof of equation~\eqref{costexpropt}}\label{app:costexpropt}
The mean field costs~\eqref{barC} can be written as
\begin{equation}
\bar C-C_0=-\frac{D^2\,h}{2}\,m\int d\theta\, \zeta^2\,\cos\theta+2\,D^2\,\int d\theta\, \zeta^2\,(\partial_\theta\ln \zeta)^2=
-\frac{D^2\,h}{2}\,m^2+2\,D^2\,\int d\theta\,(\partial_\theta \zeta)^2.
\end{equation}
We can apply partial integration to the control cost term $\int d\theta\,(\partial_\theta\zeta)^2=-\int d\theta\,\zeta\,\partial_\theta^2\zeta$. By using the Mathieu equation \ref{eqmathieuapp}, we can write $\int d\theta\,(\partial_\theta\zeta)^2=a/4+m^2\, h/2$. Equation~\eqref{costexpropt} follows from a straightforward substitution.

\section{Proofs of  the expressions for the susceptivity }\label{app:sus}
Consider an external field as defined in Eq.~\eqref{pert}. The self consistency condition~\eqref{m1} implicitly  defines the polar oder parameter $m=m(h,\epsilon)$ 
as
\begin{equation}\label{mepsilon}
\mathcal{F}(m(h,\epsilon))=m(h,\epsilon). 
\end{equation}
Upon defining $\tilde {\mathcal{F}}=m-\mathcal{F}$, Dini implicit function theorem allows to compute the susceptivity  as
\begin{equation}
\chi(h)=\left.\frac{\partial m}{\partial \epsilon}\right|_{\epsilon=0}=-\frac{\left.\frac{\partial\tilde {\mathcal{F}}}{\partial \epsilon}\right|_{\epsilon=0}}{\left.\frac{\partial\tilde {\mathcal{F}}}{\partial m}\right|_{\epsilon=0}}.
\end{equation}
though, in some cases, there is a shorter procedure. We consider the following four scenarios.
\begin{itemize}
\item $h<h_c$. In this case, we can assume that, unless there is some
  discontinuity, $\lim_{\epsilon\to 0}m(h,\epsilon)=0$. Hence,
  assuming both $\epsilon$ and $m$ small, Eq.~\eqref{mepsilon} can
  be expanded as
\begin{equation}\label{eqforubar_order3_ext}
-\frac{(h-2)\,m+2\,\epsilon/D^2}{2}+\frac{7\,(m\,h+2\,\epsilon/D^2)^3}{64}+o((m\,h+2\,\epsilon/D^2)^3)=0\,.
\end{equation}
Then we can either use Dini's theorem or simply observe that $\mathcal{F}=0$ at leading order  implies $m=2\epsilon/[D^2\,(2-h)]$ and, therefore
\begin{equation}
\chi=\frac{2}{D^2\,(h_c-h)}\,,
\end{equation}
holds exactly in this region. 

\item $h=h_c$. We can again use the ansatz $\lim_{\epsilon\to 0}m(h,\epsilon)=0$ along with Eq.~\eqref{eqforubar_order3_ext}. Then, either applying Dini's theorem, or observing that, since $h-2$ is zero exactly, leading order in $\epsilon$ must match leading order in $m$, which is $m^3$. Hence, we immediately get
\begin{equation}
m\sim 2\,\left[\frac{\epsilon}{7\,D^2}\right]^{1/3}\,,
\end{equation}
thus, at the critical point, $m(\epsilon)$ in continuous in $\epsilon$ but non differentiable. 

\item $h=h_c^+$. As in the critical regime, it remains valid that the nudge selects the direction in which the symmetry is broken since other choices would be sub-optimal. Here  $\lim_{\epsilon\to 0}m(h,\epsilon)=m(h)>0$, however, as long as $h$ is close to $h_c$, we can still use expansion~\eqref{eqforubar_order3_ext}. From Eq.~\eqref{m_vs_h_opt} we can write $m(h)=\sqrt{(4/7)\,(h-2)}+\delta m$, with $\delta m\ll 1$ and match leading order of $\delta m$ and $\epsilon$. As a result, we get
\begin{equation}
\chi\sim \frac{1}{D^2\,(h_c-h)}\,.
\end{equation}

\item $h\gg h_c$. Using Eq.~\eqref{u_and_h_gauss}, the self consistency equation \eqref{m1} becomes
\begin{equation}
m-e^{-\frac{1}{2\,\sqrt{h\,m+2\,\epsilon/D^2}}}=0\,,
\end{equation}
and a straightforward application of  Dini's theorem yields
\begin{equation}
\chi\sim \frac{1}{2\,D^2\,h^{3/2}\,{m}^{1/2}}\,.
\end{equation}
\end{itemize}

\section{Modified Bessel function of the first kind}\label{app:Bessel}
Modified Bessel functions of the first kind are non decreasing solution of the modified Bessel equation:
\begin{equation}\label{bess}
z^2 \,I_n''+z\, I_n'-(z^2+n^2)\,I_n=0
\end{equation}
We report a few identities we have used in the main text
\begin{align}
I_0(z)&=\int_{-\pi}^\pi d\theta \,e^{z\cos\theta}\label{bess0}\\
I_0'(z)&=I_1(z)=\int_{-\pi}^\pi d\theta \,\cos\theta\,e^{z\cos\theta}\label{bess1}\\
I_0''(z)&=I_0(z)-\frac{1}{z}\,I_1(z)=\int_{-\pi}^\pi d\theta \,\cos^2\theta\,e^{z\cos\theta}\label{bess2}
\end{align}
For further details, see~\cite{gradshteyn2014table}.

\section{Best sinusoidal control near the critical point}\label{app:SinCrit}
To obtain some analytical insights into the critical behavior of the
best sinusoidal model, we follow the same logic as for the optimal
model (\ref{app:OptiCrit}). The main difference is that, instead of
expanding for small $q$, we expand for small $K$. In the end, the
procedures appear to be equivalent as they are both expansion in
$\sqrt{h-h_c}$.  First, we can expand~\eqref{VonMises} and get
\begin{multline}\label{smallVonMises}
  \rho_s=\frac{1}{2\,\pi}\left(1+K\cos\theta+\frac{K^2}{4}(2\,\cos^2\theta-1)
 \right.\\
 \left.
  +\frac{K^3}{12}(2\,\cos^3\theta-3\cos\theta)+\frac{K^4}{192}(9-24\,\cos^2\theta+8\,\cos^4\theta)\right)+o(K^4)
\end{multline}
Then, by using Eqs.~\eqref{mK} and~\eqref{smallVonMises}, the average total cost~\eqref{costSin} can be written explicitly as
\begin{equation}\label{Vic_cost_small_k}
\bar C_s-C_0=D^2\left[\left(\frac{1}{4}-\frac{h}{8}\right)\,K^2+\frac{h-1}{32}\,K^4\right]+o(K^4)
\end{equation}
A positive $K=K_\star$ which minimizes the previous expression exists only for $h> h_c=2$ and its asymptotic expression near $h_c$ is
\begin{equation}\label{K_vs_h_Vic_app}
K_\star\sim \sqrt{2\,(h-2)}\,,
\end{equation}
which plugged into Eq.~\eqref{mK} yields
\begin{equation}\label{u_vs_h_Vic_app}
m\sim\sqrt{\frac{1}{2}(h-h_c)}\,.
\end{equation}
It follows from Eqs.~\eqref{K_vs_h_Vic_app} and~\eqref{Vic_cost_small_k} that
\begin{equation}\label{csin}
\bar C_s-C_0=-\frac{D^2}{8}\,(h-h_c)^2+o((h-h_c)^2)\,.
\end{equation}

\newpage

\section*{References}
%\bibliographystyle{iopart-num}
%\bibliography{biblioflock.bib}

\begin{thebibliography}{10}
\expandafter\ifx\csname url\endcsname\relax
  \def\url#1{{\tt #1}}\fi
\expandafter\ifx\csname urlprefix\endcsname\relax\def\urlprefix{URL }\fi
\providecommand{\eprint}[2][]{\url{#2}}

\bibitem{okubo1986dynamical}
Okubo A 1986 {\em Adv. Biophys.\/} {\bf 22} 1--94

\bibitem{vicsek2012collective}
Vicsek T and Zafeiris A 2012 {\em Phys. Rep.\/} {\bf 517} 71--140

\bibitem{wolgemuth2008collective}
Wolgemuth C~W 2008 {\em Biophys. J.\/} {\bf 95} 1564--1574

\bibitem{sullivan1981insect}
Sullivan R~T 1981 {\em Fla. Entomol.\/} {\bf 64} 44--65

\bibitem{cavagna2017dynamic}
Cavagna A, Conti D, Creato C, Del~Castello L, Giardina I, Grigera T~S, Melillo
  S, Parisi L and Viale M 2017 {\em Nat. Phys.\/} {\bf 13} 914--918

\bibitem{cavagna2018physics}
Cavagna A, Giardina I and Grigera T~S 2018 {\em Phys. Rep.\/} {\bf 728} 1--62

\bibitem{attanasi2014information}
Attanasi A, Cavagna A, Del~Castello L, Giardina I, Grigera T~S, Jeli{\'c} A,
  Melillo S, Parisi L, Pohl O, Shen E and Viale M 2014 {\em Nature Phys.\/}
  {\bf 10} 691--696

\bibitem{pitcher1983heuristic}
Pitcher T~J 1983 {\em Anim. Behav.\/} {\bf 31} 611--613

\bibitem{pavlov2000patterns}
Pavlov D and Kasumyan A 2000 {\em J. Ichthyol.\/} {\bf 40} S163

\bibitem{reynolds1987flocks}
Reynolds C~W 1987 Flocks, herds and schools: A distributed behavioral model
  {\em Proceedings of the 14th annual conference on Computer graphics and
  interactive techniques\/} pp 25--34

\bibitem{vicsek1995novel}
Vicsek T, Czir{\'o}k A, Ben-Jacob E, Cohen I and Shochet O 1995 {\em Phy. Rev.
  Lett.\/} {\bf 75} 1226

\bibitem{turgut2008self}
Turgut A~E, {\c{C}}elikkanat H, G{\"o}k{\c{c}}e F and {\c{S}}ahin E 2008 {\em
  Swarm. Intell.\/} {\bf 2} 97--120

\bibitem{viragh2014flocking}
Vir{\'a}gh C, V{\'a}s{\'a}rhelyi G, Tarcai N, Sz{\"o}r{\'e}nyi T, Somorjai G,
  Nepusz T and Vicsek T 2014 {\em Bioinspir. Biomim.\/} {\bf 9} 025012

\bibitem{todorov2009efficient}
Todorov E 2009 {\em Proc. Natl. Acad. Sci. U.S.A.\/} {\bf 106} 11478--11483

\bibitem{lasry2007mean}
Lasry J~M and Lions P~L 2007 {\em Japanese J. Math.\/} {\bf 2} 229--260

\bibitem{ullmo2019quadratic}
Ullmo D, Swiecicki I and Gobron T 2019 {\em Phys. Rep.\/} {\bf 799} 1--35

\bibitem{pezzotta2018chemotaxis}
Pezzotta A, Adorisio M and Celani A 2018 {\em Phys. Rev. E\/} {\bf 98} 042401

\bibitem{hongler2020mean}
Hongler M~O 2020 {\em J. Dyn. Games\/} {\bf 7} 1

\bibitem{durve2020learning}
Durve M, Peruani F and Celani A 2020 {\em Phys. Rev. E\/} {\bf 102} 012601

\bibitem{romanczuk2012active}
Romanczuk P, B{\"a}r M, Ebeling W, Lindner B and Schimansky-Geier L 2012 {\em
  Europ. Phys. J. Spec. Top.\/} {\bf 202} 1--162

\bibitem{bechinger2016active}
Bechinger C, Di~Leonardo R, L{\"o}wen H, Reichhardt C, Volpe G and Volpe G 2016
  {\em Rev. Mod. Phys.\/} {\bf 88} 045006

\bibitem{mora2011biological}
Mora T and Bialek W 2011 {\em J. Stat. Phys.\/} {\bf 144} 268--302

\bibitem{farrell2012pattern}
Farrell F~D~C, Marchetti M~C, Marenduzzo D and Tailleur J 2012 {\em Phys. Rev.
  Lett.\/} {\bf 108} 248101

\bibitem{chepizhko2021revisiting}
Chepizhko O, Saintillan D and Peruani F 2021 {\em Soft Matter\/} {\bf 17}
  3113--3120

\bibitem{peruani2008mean}
Peruani F, Deutsch A and B{\"a}r M 2008 {\em Europ. Phys. J. Spec. Top.\/} {\bf
  157} 111--122

\bibitem{chepizhko2009kinetic}
Chepizhko A and Kulinskii V 2009 The kinetic regime of the vicsek model {\em
  AIP Conf. Proc.\/} vol 1198 pp 25--33

\bibitem{chepizhko2010relation}
Chepizhko A and Kulinskii V 2010 {\em Physica A\/} {\bf 389} 5347--5352

\bibitem{howard1972risk}
Howard R~A and Matheson J~E 1972 {\em Manage. Sci.\/} {\bf 18} 356--369

\bibitem{dvijotham2011unified}
Dvijotham K and Todorov E 2011 A unified theory of linearly solvable optimal
  control {\em Proceedings of the 27th Conference on Uncertainty in Artificial
  Intelligence (UAI 2011)\/} vol~1 pp 25--34

\bibitem{pontryagin2018mathematical}
Pontryagin L~S 2018 {\em Mathematical theory of optimal processes\/}
  (Routledge)

\bibitem{brimacombe2020computation}
Brimacombe C, Corless R~M and Zamir M 2020 {\em arXiv preprint
  arXiv:2008.01812\/}

\bibitem{gutierrez2003mathieu}
Guti{\'e}rrez-Vega J~C, Rodr{\i}guez-Dagnino R, Meneses-Nava M and
  Ch{\'a}vez-Cerda S 2003 {\em Am. J. Phys.\/} {\bf 71} 233--242

\bibitem{bialek2012statistical}
Bialek W, Cavagna A, Giardina I, Mora T, Silvestri E, Viale M and Walczak A~M
  2012 {\em Proc. Natl. Acad. Sci. U.S.A.\/} {\bf 109} 4786--4791

\bibitem{kuramoto1975self}
Kuramoto Y 1975 Self-entrainment of a population of coupled non-linear
  oscillators {\em International symposium on mathematical problems in
  theoretical physics\/} (Springer) pp 420--422

\bibitem{acebron2005kuramoto}
Acebr{\'o}n J~A, Bonilla L~L, Vicente C~J~P, Ritort F and Spigler R 2005 {\em
  Rev. Mod. Phys\/} {\bf 77} 137

\bibitem{ginelli2016physics}
Ginelli F 2016 {\em Eur. Phys. J. Spec. Top.\/} {\bf 225} 2099--2117

\bibitem{pedley1987orientation}
Pedley T~J and Kessler J 1987 {\em Proc. R. Soc. B\/} {\bf 231} 47--70

\bibitem{cencini2016centripetal}
Cencini M, Franchino M, Santamaria F and Boffetta G 2016 {\em J. Theor.
  Biol.\/} {\bf 399} 62--70

\bibitem{mardia2009directional}
Mardia K~V and Jupp P~E 2009 {\em Directional statistics\/} vol 494 (John Wiley
  \& Sons)

\bibitem{gradshteyn2014table}
Gradshteyn I~S and Ryzhik I~M 2014 {\em Table of integrals, series, and
  products\/} (Academic press)

\end{thebibliography}

\providecommand{\newblock}{}

\end{document}